\documentclass[twocolumn,showpacs,preprintnumbers,amsmath,amssymb]{revtex4}

\usepackage{graphicx}
\usepackage{dcolumn}
\usepackage{bm}

\begin{document}

\preprint{APS/123-QED}

\title{Circuit analysis of quantum measurement}

\author{Yuji Kurotani$^1$}
\author{Masahito Ueda$^{1,2}$}
\affiliation{$^1$ Department of Physics, Tokyo Institute of Technology, Tokyo 152-8551, Japan\\
$^2$ Macroscopic Quantum Control Project, ERATO, JST, 2-11-16 Yayoi, Bunkyo-ku, Tokyo 113-8656, Japan}

\date{\today}

\begin{abstract}
We develop a circuit theory that enables us to analyze quantum measurements on a two-level system and on a continuous-variable system on an equal footing. As a measurement scheme applicable to both systems, we discuss a swapping state measurement which exchanges quantum states between the system and the measuring apparatus before the apparatus meter is read out.
This swapping state measurement has an advantage in gravitational-wave detection over contractive state measurement in that the postmeasurement state of the system can be set to a prescribed one, regardless of the outcome of the measurement. 
\end{abstract}



\maketitle

\section{Introduction}
\label{sec:Introduction}

Von Neumann has established a general framework for quantum measurement by postulating that any indirect measurement can be decomposed into two distinct parts~\cite{von_Neumann}.
The first part involves unitary transformation that transfers the information of interest from the measured system to the measuring apparatus.  
The second part involves nonunitary transformation that achieves the realization of a particular outcome of the measurement by the projection postulate.
The adjective indirect implies that the projection postulate is applied not to the measured system but to the measuring apparatus.
As an illustrative example, von Neumann constructed a model for the position measurement of a massive particle which is referred to as von Neumann measurement (VNM).

VNM satisfies Heisenberg's noise-disturbance uncertainty relation~\cite{Heisenberg,Ozawa2003} and sets the standard quantum limit (SQL)~\cite{Braginsky, Caves_et_al} on the accuracy of repeated position measurements.
This noise-disturbance uncertainty relation refers to a trade-off relation between the noise added to a system's observable and the disturbance generated in the conjugate observable due to the back-action of the measurement.
The measurement \textit{noise} of observable $\hat A$ is defined by the difference between the probability distribution of obtaining a measurement outcome and the corresponding probability distribution calculated from the premeasurement state of the system according to Born's probability axiom~\cite{Ozawa2003}. If there is no difference for an arbitrary input state of the system, then the measurement of observable $\hat A$ is said to be \textit{noiseless}. 
In 1983, Yuen suggested the possibility of a contractive state measurement (CSM) in which Heisenberg's noise-disturbance uncertainty relation is violated and the SQL is surpassed~\cite{Yuen}.
Ozawa subsequently presented a concrete model that vindicates Yuen's conjecture~\cite{Ozawa, Ozawa_book}.

In the present paper we discuss a swapping state measurement (SSM) in which quantum states are exchanged between the measured system and the measuring apparatus before the projection postulate is applied to the apparatus~\cite{Yuen_SSM}.
While VNM, CSM, and SSM are all noiseless, SSM is unique in that it is noiseless for an arbitrary observable.
As discussed in Sec.~\ref{sec:Gravitation}, SSM has the advantage over CSM in gravitational-wave detection in respect of experimental implementation of the measuring apparatus.
SSM is shown to be described with three SUM gates, where the SUM gate is the continuous-variable counterpart of the controlled-NOT (CNOT) gate~\cite{Gottesman, Bartlett, Braunstein}.
In contrast, VNM and CSM can be respectively described with one and two SUM gates~\cite{Ozawa2001}.
We discuss the two-level counterparts of these three models and develop a quantum circuit theory to establish complete parallelism between continuous-variable systems and two-level systems.

This paper is organized as follows. 
In Sec.~\ref{sec:spin}, we discuss noiseless measurements on a two-level system,  and describe the two-level counterparts of VNM, CSM, and SSM in terms of CNOT gates. 
In Sec.~\ref{sec:CV}, we discuss three noiseless measurements on a continuous-variable system in terms of SUM gates.
In Sec.~\ref{sec:IHUG}, we construct a concrete Hamiltonian for each model and express the corresponding unitary transformation in terms of single-qubit rotations and the SWAP operator.
We also decompose unitary transformation in a continuous-variable system into a phase-shift operator, squeezing operator and beam-splitter operator to enable experimental implementation of the models.
In Sec.~\ref{sec:Discussions}, we discuss basic properties of the three noiseless measurements and compare SSM and CSM for use in gravitational-wave detection.
In Sec.~\ref{sec:Conclusion}, we summarize the main results of this paper.

Throughout this paper we refer to a measured system and a measuring apparatus simply as \textit{system} and \textit{probe}, respectively.
We shall, as with VNM~\cite{von_Neumann} and CSM~\cite{Ozawa, Ozawa_book}, ignore the free parts of the Hamiltonian and focus only on the interaction between the system and the probe.

\section{Quantum measurement in a two-level system}
\label{sec:spin}

We consider a situation in which the system and the probe are both two-level systems with their initial states given by
\begin{align}
|\psi \rangle_{\rm s} &=a |+ \rangle_{\rm s} +b |- \rangle_{\rm s}, \label{input_s} \\
|\phi \rangle_{\rm p} &=c |+ \rangle_{\rm p} +d |- \rangle_{\rm p}, \label{input_p}
\end{align}
where $a, b, c, d$ are complex numbers, and $|\pm \rangle$ stand for the eigenstates of the Pauli $\hat \sigma^z$ operator with 
eigenvalues $s^z=\pm1$.
We assume that both system state $|\psi \rangle_{\rm s}$ and probe state $|\phi \rangle_{\rm p}$ are normalized to unity.

\subsection{CNOT measurement}

\begin{figure}[htbp]
\begin{center}
\includegraphics[width=4.0cm,clip]{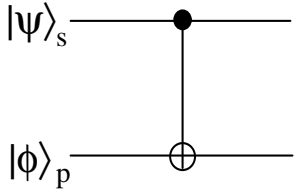}
\caption{CNOT circuit. }
\label{fig:CNOT}
\end{center}
\end{figure}

Let us first consider a CNOT measurement.
A unitary operation of CNOT measurement is performed by the CNOT circuit~\cite{Feynman, Deutsch} illustrated in Fig.~\ref{fig:CNOT}.
If the input state of the system is $|+ \rangle_{\rm s}$, the output state of the probe is the same as its input state ($|\pm \rangle_{\rm p}\rightarrow|\pm \rangle_{\rm p}$). If it is $|- \rangle_{\rm s}$, the parity of the output state of the probe is reversed ($|\pm \rangle_{\rm p}\rightarrow|\mp \rangle_{\rm p}$).

Output state $|\Psi \rangle_{\rm s+p}^{\rm CNOT}$ of the CNOT circuit for input states (\ref{input_s}) and (\ref{input_p}) is given by
\begin{equation}
\begin{split}
&|\Psi \rangle_{\rm s+p}^{\rm CNOT}=\hat U^{\text{CNOT}}_{\rm{s} \text{-} \rm p}|\psi\rangle_{\rm s}|\phi\rangle_{\rm p}\\
=&\ \left(ac |+ \rangle_{\rm s} + bd |- \rangle_{\rm s}\right) |+ \rangle_{\rm p}+\left(ad |+ \rangle_{\rm s} +bc |- \rangle_{\rm s}\right) |- \rangle_{\rm p}.
\end{split}
\end{equation}
Suppose that we perform a projection measurement of $\hat \sigma^z$ on the output state of the probe. Then the probabilities of obtaining outcomes $s^z_{\rm p}=\pm1$ are given by
\begin{eqnarray}
\text{P}^{\rm CNOT}\left[s^z_{\rm p}=1\right] &=& |ac|^2+|bd|^2 ,  \nonumber \\
\text{P}^{\rm CNOT}\left[s^z_{\rm p}=-1 \right] &=& |ad|^2+|bc|^2 .
\label{eq:CNOT_prob}
\end{eqnarray}
When $c=1$ and $d=0$, Eq.~\eqref{eq:CNOT_prob} reduces to
\begin{eqnarray}
& & \text{P}^{\rm CNOT}\left[s^z_{\rm p}=1\right] = |a|^2,  \nonumber\\
& & \text{P}^{\rm CNOT}\left[s^z_{\rm p}=-1 \right] = |b|^2.
\label{eq:CNOT_prob_Born}
\end{eqnarray}
These probabilities coincide with $|{}_{\rm s}\langle \pm|\psi\rangle_{\rm s}|^2$ which can be expected from Born's probability axiom.
The CNOT measurement is therefore a noiseless measurement of system's $\hat \sigma^z$ if the probe state is properly chosen, i.e., if $c=1$ and $d=0$.
The postmeasurement states of the system $|\psi'\rangle_{\rm s}$ for outcomes $s^z_{\rm p}=\pm 1$ are then given by
\begin{equation}
|\psi'\rangle_{\rm s}^{\rm CNOT}=|\pm \rangle_{\rm s}.
\label{eq:CNOT_final}
\end{equation}
We note that the postmeasurement states of the CNOT measurement depend only on the outcome of the measurement.

The indirect measurement can be operationally described by a set of \textit{measurement operators} $\{ \hat M_m\}$ that directly act on the measured system~\cite{Gordon_Louisell, Davies_Lewis} and satisfy completeness relation $\sum_{m}\hat M_m^{\dagger} \hat M_m= \hat I$.
Let the premeasurement state of the system be $|\psi\rangle$.
The probability of outcome $m$ being found is then given by 
\begin{equation}
P(m)=\langle \psi | \hat M_m^{\dagger} \hat M_m| \psi \rangle, 
\label{probability}
\end{equation}
and corresponding postmeasurement state $|\psi'\rangle_m$ of the system is given by 
\begin{equation}
|\psi'\rangle_m= \frac{\hat M_m|\psi\rangle}{\sqrt{\langle \psi |\hat M_m^{\dagger} \hat M_m| \psi \rangle}}.
\label{projection}
\end{equation}
It follows from Eqs.~(\ref{eq:CNOT_prob_Born})-(\ref{projection}) that the CNOT measurement with $c=1$ and $d=0$ can be characterized with measurement operators 
\begin{equation}
\hat M_{\pm}^{\rm CNOT}=|\pm \rangle_{\rm s} {}_{\rm s}\langle \pm|.
\label{M_CNOT}
\end{equation}

\subsection{Double CNOT measurement}

\begin{figure}[htbp]
\begin{center}
\includegraphics[width=4cm,clip]{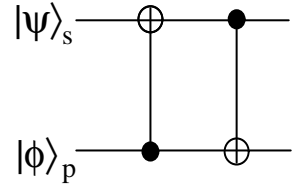}
\caption{Double CNOT (DCNOT) circuit.}
\label{fig:Double_CNOT}
\end{center}
\end{figure}

Let us next consider a double CNOT (DCNOT) measurement.
A unitary operation of the DCNOT measurement is represented by the DCNOT circuit illustrated in Fig.~\ref{fig:Double_CNOT}.

Output state $|\Psi \rangle_{\rm s+p}^{\rm DCNOT}$ of the DCNOT circuit for input states (\ref{input_s}) and (\ref{input_p}) is given by
\begin{eqnarray}
& & |\Psi \rangle_{\rm s+p}^{\rm DCNOT}
=\hat U^{\text{DCNOT}}_{\rm s\text{-} \rm p}|\psi\rangle_{\rm s}|\phi\rangle_{\rm p}
\nonumber \\
&=&\left(c |+ \rangle_{\rm s} + d |- \rangle_{\rm s}\right) a|+ \rangle_{\rm p} 
+\left(d |+ \rangle_{\rm s} +c |- \rangle_{\rm s}\right) b|- \rangle_{\rm p}.
\nonumber
\\
\end{eqnarray}
Suppose that we perform a projection measurement of $\hat \sigma^z$ on the output state of the probe. Then the probabilities of obtaining outcomes $s^z_{\rm p}=\pm1$ are given by
\begin{eqnarray}
& & \text{P}^{\rm DCNOT}\left[s^z_{\rm p}=1 \right] = |a|^2, \nonumber\\
& & \text{P}^{\rm DCNOT}\left[s^z_{\rm p}=-1\right] = |b|^2.
\label{eq:double_prob}
\end{eqnarray}
These probabilities are independent of the state of the probe, unlike the case of the CNOT measurement in Eq.~\eqref{eq:CNOT_prob}.
The corresponding postmeasurement states of the system are given by 
\begin{eqnarray}
& |\psi'_{+} \rangle_{\rm s}^{\rm DCNOT}
=c |+ \rangle_{\rm s} + d |- \rangle_{\rm s}  
\ \ & {\rm for} \ s^z_{\rm p}=1,
\nonumber\\
& |\psi'_{-} \rangle_{\rm s}^{\rm DCNOT}
=d |+ \rangle_{\rm s} + c |- \rangle_{\rm s} 
\ \ & {\rm for} \ s^z_{\rm p}=-1.
\label{eq:double_post}
\end{eqnarray}
The postmeasurement state now depends not only on the outcome of the measurement but also on the state of the probe. 
It follows from Eqs.~(\ref{probability}), (\ref{projection}), (\ref{eq:double_prob}) and (\ref{eq:double_post}) that the DCNOT measurement is characterized with measurement operators 
\begin{eqnarray}
\hat M_{\pm}^{\rm DCNOT}
=|\psi'_{\pm} \rangle_{\rm s}^{\rm DCNOT} {}_{\rm s} \langle \pm |.
\label{M_double}
\end{eqnarray}

\subsection{Swapping state measurement}

\begin{figure}[htbp]
\begin{center}
\includegraphics[width=4.5cm,clip]{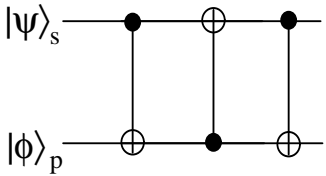}
\caption{SWAP circuit.}
\label{fig:spin_swap}
\end{center}
\end{figure}

Let us now consider the swapping state measurement (SSM).
Swap operator $\hat U^{\text{SWAP}}$ exchanges quantum states between the system and the probe and is defined by~\cite{Feynman}
\begin{equation}
\hat U^{\text{SWAP}}(|\psi\rangle_{\rm s} |\phi\rangle_{\rm p})
=|\phi\rangle_{\rm s} |\psi\rangle_{\rm p}.
\label{USWAP}
\end{equation}
A unitary circuit of SSM can be expressed in terms of three CNOT gates as illustrated in Fig.~\ref{fig:spin_swap}.

Output state $|\Psi \rangle_{\rm s+p}^{\rm SWAP}$ of the SWAP circuit for input states (\ref{input_s}) and (\ref{input_p}) is given by
\begin{equation}
|\Psi \rangle_{\rm s+p}^{\rm SWAP}=(c |+ \rangle_{\rm s}+ d|- \rangle_{\rm s}) (a |+ \rangle_{\rm p}+b |-\rangle_{\rm p}).
\label{SWAPoutput}
\end{equation}
We note that this is not an entangled state but a product state.
Suppose that we perform the projection measurement of $\hat \sigma^z$ on the output probe state. Then the probabilities of obtaining outcomes $s^z_{\rm p}=\pm1$ are given by
\begin{eqnarray}
& & \text{P}^{\rm SWAP}\left[s^z_{\rm p}=1\right] = |a|^2, \noindent \\
& & \text{P}^{\rm SWAP}\left[s^z_{\rm p}=-1\right] = |b|^2,
\label{eq:swap_prob}
\end{eqnarray}
which are independent of the state of the probe as in the case of the DCNOT measurement in Eq.~\eqref{eq:double_prob}.
Postmeasurement state of the system $|\psi'\rangle_{\rm s}$ is given for both $s^z_{\rm p}=1$ and $s^z_{\rm p}=-1$ by
\begin{equation}
|\psi'\rangle_{\rm s}^{\rm SWAP}= c |+ \rangle_{\rm s} + d |- \rangle_{\rm s},
\label{eq:swap_post}
\end{equation}
indicating that the postmeasurement state of the system is independent of the measurement outcome and is identical to the state of the probe. 
We may use this property to designate the postmeasurement state of the system by preparing the initial probe state.
It follows from Eqs.~\eqref{probability}, \eqref{projection}, \eqref{SWAPoutput}, and \eqref{eq:swap_post} that SSM can be characterized by measurement operators 
\begin{equation}
\hat M_{\pm}^{\rm SWAP}=|\phi \rangle_{\rm s} {}_{\rm s} \langle \pm |.
\label{M_SWAP}
\end{equation}

While we consider here the Pauli $\hat \sigma^z$ measurement,
the present noiseless measurement scheme can be applied to an arbitrary observable by using a similar SWAP circuit because the SWAP circuit exchanges quantum states between the system and the probe.

\section{Quantum Measurement in a Continuous-Variable System}
\label{sec:CV}

We consider a situation in which both the \textit{system} and the \textit{probe} are one-dimensional, having canonically conjugate observables $\hat x$, $\hat p_x$ $\left([\hat x, \hat p_x]=i \right)$ and $\hat y$, $\hat p_y$ $\left([\hat y, \hat p_y]=i \right)$, respectively.
Let the initial wave functions of the system and the probe be given by $\langle x| \psi\rangle =\psi(x)$ and $\langle y| \phi\rangle=\phi(y)$, respectively, which we assume to be normalized to unity.

\subsection{General unitary transformation}

Let us consider unitary transformation $\hat U$ such that
\begin{equation}
\hat U(a,b,c,d)\psi(x)\phi(y)=\psi(ax+by)\phi(cx+dy),
\label{eq:Trans}
\end{equation}
where $a,b,c$ and $d$ are real numbers.
The unitarity condition of $U$ implies that 
\begin{equation}
\int \big| \psi(x) \phi(y) \big|^2 dxdy= \int \big| \psi (ax+by)\phi (cx+dy)\big|^2 dxdy,
\label{eq:norm2}
\end{equation}
which is satisfied if and only if the following condition is met:
\begin{equation}
ad-bc=\pm 1,
\end{equation}
where the minus sign implies that an odd parity inversion is involved in the unitary transformation (see Eqs.~\eqref{parity} and \eqref{eq:det} below).

We define von Neumann unitary operators $\hat V_{xp_y}$ and $\hat V_{yp_x}$ as
\begin{eqnarray}
& & \hat V_{xp_y}(\alpha )\equiv\exp(-i\alpha \hat x \hat p_y), 
\label{VN1} \\
& & \hat V_{yp_x}(\alpha )\equiv\exp(-i\alpha \hat y \hat p_x),
\label{VN2}
\end{eqnarray}
where $\alpha$ is a real number.
For $\alpha=1$, we refer to the von Neumann operator as a SUM gate which is known as a continuous-variable analog of a CNOT gate~\cite{Gottesman, Bartlett, Braunstein}. 
We also define parity inversion operator $\hat \Pi_y$ of the probe as
\begin{equation}
\hat \Pi_y\phi(y)=\phi(-y).
\label{parity}
\end{equation}

\begin{figure}[htbp]
\begin{center}
\includegraphics[height=3.3cm,clip]{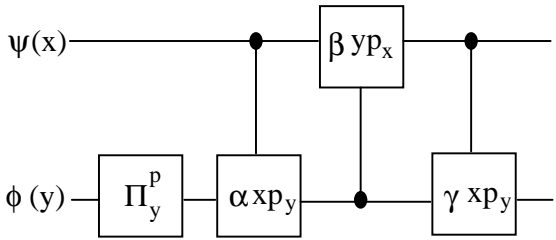}
\caption{Circuit representation of general unitary transformation $U$ in Eq.~\eqref{eq:Trans} in terms of three basic quantum gates  $\hat V_{yp_x}$, $\hat V_{xp_y}$ and $\hat \Pi_y$ (see Eq.~\eqref{eq:decompose} for the mathematical expression).}
\label{fig:circuit}
\end{center}
\end{figure}

We now show that unitary operator $\hat U$ can be decomposed into a product of $\hat V_{xp_y}$, $\hat V_{yp_x}$ and $\hat \Pi_y$.
In fact, for $p=0$ or $1$, we have
\begin{equation}
\begin{split}
&\hat V_{xp_y}(\gamma )\hat V_{yp_x}(\beta )\hat V_{xp_y}(\alpha )\hat \Pi_y^p \psi\left(x \right)\phi\left(y \right) \\
=&\hat V_{yp_x}(\gamma )\hat V_{xp_y}(\beta )\psi\left(x \right)\phi\left[(-1)^p(y-\alpha x) \right ] \\
=&\hat V_{yp_x}(\gamma )\psi (x-\beta y)\phi\left((-1)^p\left[y-\alpha (x-\beta y)\right] \right ) \\
=&\psi\left[(1+\beta \gamma ) x-\beta y\right] \\
&
\!\!\!\!\times \!\!\!\!\!\!
\hspace{0.3cm} \phi\left[(-1)^{p+1} (\alpha+\gamma+\alpha\beta \gamma) x
+\!(-1)^p(1+\alpha \beta) y\right].
\end{split}
\label{eq:decompose_t}
\end{equation}
The last term is cast into the form of $\psi(ax+by)\phi(cx+dy)$ if the conditions
\begin{eqnarray}
a&=&1+ \beta \gamma, \nonumber \\
b&=&-\beta,          \nonumber \\
c&=&(-1)^{p+1} (\alpha+\gamma+\alpha\beta \gamma), \nonumber \\
d&=&(-1)^p(1+\alpha \beta),
\label{eq:abcd}
\end{eqnarray}
are met.
It follows from Eq.~\eqref{eq:abcd} that parameters $a,b,c$ and $d$ satisfy
\begin{equation}
ad-bc=(-1)^p,
\label{eq:det}
\end{equation}
and that unitary operator $U(a,b,c,d)$ can be decomposed into
\begin{equation}
\hat U(a,b,c,d)=\hat V_{xp_y}(\gamma )\hat V_{yp_x}(\beta )\hat V_{xp_y}(\alpha )\hat \Pi_y^p.
\label{eq:decompose}
\end{equation}
We thus find that unitary operator $\hat U$ can be decomposed into three basic quantum gates $\hat V_{yp_x}$, $\hat V_{xp_y}$ and $\hat \Pi_y$ as illustrated in Fig.~\ref{fig:circuit}.

\subsection{Von Neumann position measurement}
\label{sec:von Neumann}

\begin{figure}[htbp]
\begin{center}
\includegraphics[height=3.0cm,clip]{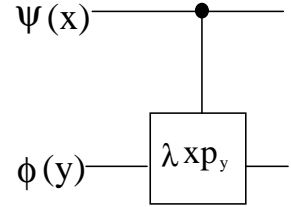}
\caption{VNM circuit representation of a generalized von Neumann's position measurement.}
\label{fig:neumann}
\end{center}
\end{figure}

Von Neumann measurement (VNM) is a prototypical indirect measurement on a continuous-variable system~\cite{von_Neumann}.
Unitary transformation $\hat U$ corresponding to von Neumann's position measurement is given by Eq.~\eqref{eq:decompose} with $(a,b,c,d)=(1,0,-1,1)$.
We consider here a generalized unitary transformation characterized by $(a,b,c,d)=(1,0,-\lambda,1)$, where parameter $\lambda$ is a real positive number.
We shall refer to $\lambda$ as a scaling parameter because it leads to a scale transformation of the probability distribution as discussed next.
We find from Eq.~\eqref{eq:abcd} that the parameters characterizing the circuit representation are $(p,\alpha,\beta,\gamma)=(0,0,0,\lambda)$ which gives the VNM circuit as illustrated in Fig.~\ref{fig:neumann}.

We assume that the input state for VNM is $\psi(x)\phi(y)$, where $\psi(x)$ and $\phi(y)$ are the respective wave functions of the system and the probe.
The output state of the VNM circuit is then given by 
\begin{equation}
\hat U(1,0,-\lambda,1) \psi(x) \phi (y)=\psi(x)\phi(y-\lambda x).
\label{VNMoutput}
\end{equation}
Suppose that we perform position measurement on the probe for the state represented by Eq.~\eqref{VNMoutput}.
The probability of finding an outcome in $a\leq y\leq a+da$ is then given by
\begin{equation}
\text{P} \{ a\leq y \leq a+da \} =da \int |\psi (x)|^2 |\phi(a- \lambda x)|^2 dx.
\label{eq:Neumann_prob}
\end{equation}
Hence, we find that as the probe's initial probability distribution $|\phi(y)|^2$ approaches $\delta(y)$, Eq.~\eqref{eq:Neumann_prob} reduces to
\begin{equation}
\text{P} \{ a\leq y\leq a+da \}=\frac{1}{\lambda}\left|\psi\left(\frac{a}{\lambda}\right)\right|^2da.
\label{eq:Neumann_ideal_prob}
\end{equation}
We note that probability distribution~\eqref{eq:Neumann_ideal_prob} can be obtained from distribution $|\psi (a)|^2$ by reducing its magnitude by a factor of $\lambda$ and expanding the scale of the argument by the same factor.
The scaling parameter thus leads to a scale transformation of the probability distribution.
In the special case of $\lambda=1$, the probability distribution reduces to $|\psi (a)|^2$ which is to be expected from Born's probability axiom; this measurement is therefore a noiseless measurement of position $x$ of the system.

Returning to the case of Eq.~\eqref{eq:Neumann_prob}.
Corresponding postmeasurement system state $\psi'_a(x)$ is given by
\begin{equation}
\psi'_a(x)= \frac{\psi(x)\phi(a-\lambda x)}{(\int |\psi (x)|^2 |\phi(a- \lambda x)|^2 dx)^{1/2}}.
\label{eq:VNM_post}
\end{equation}
This wave function approaches $\left[\delta(x-a/\lambda)\right]^{1/2}$
as the probe's initial probability distribution $|\phi(y)|^2$ approaches $\delta(y)$.
This implies that the postmeasurement state approaches an eigenstate of the position operator as the position measurement of the probe becomes noiseless. 
In particular, when $\lambda=1$ and $|\phi(y)|^2=\delta(y)$, VNM is characterized by measurement operator $\hat M_a=|a \rangle \langle a|$.
Comparing this with Eq.~\eqref{M_CNOT}, we find close similarity between VNM and the CNOT measurement.

The scaling parameter can be used to improve the precision of VNM~\cite{Yurke}.
As an example, let us consider the situation in which the initial probability distribution of the system is known to be delta function $|\psi(x)|^2=\delta(x-\alpha)$ but its location $\alpha$ is unknown.
It follows from Eqs.~\eqref{eq:Neumann_prob} and \eqref{eq:VNM_post} that the probability distribution for the measurement outcome of the probe is given by $\text{P} \{ a\leq y \leq a+da \} =|\phi(a- \lambda \alpha)|^2 da$ and that the corresponding postmeasurement state of the system is given by $|\psi'_a(x)|^2=\delta(x-\alpha)$.
Hence, we may regard this measurement as a quantum nondemolition measurement of the position because its probability distribution does not change before and after the measurement.
By properly choosing the origin of the probe's coordinate, we can always set $\int y |\phi(y)|^2 dy=0$. 
The signal-to-noise (S/N) ratio of the measurement is then given by
\begin{equation}
\frac{\rm S}{\rm N}\equiv \frac{\left(\int aP(a) da\right)^2}{\int a^2P(a) da-\left(\int aP(a) da\right)^2}
=\frac{\lambda^2\alpha^2}{d^2},
\label{S/N}
\end{equation}
where $P(a)=|\phi(a- \lambda \alpha)|^2$ and $d\equiv\sqrt{\int y^2 |\phi(y)|^2 dy}$.
Equation~\eqref{S/N} shows that, even if the signal-to-noise ratio of the probe is low, we can measure the position of the system to the desired precision by choosing sufficiently large scaling parameter $\lambda$.

\subsection{Contractive state measurement}

\begin{figure}[htbp]
\begin{center}
\includegraphics[height=3.5cm,clip]{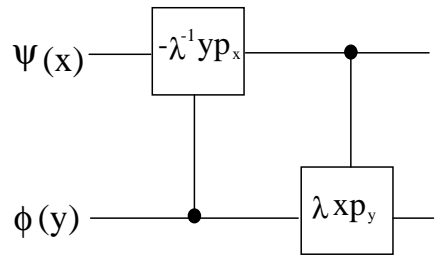}
\caption{Circuit representation of generalized contractive state measurement.}
\label{fig:CSM}
\end{center}
\end{figure}

The notion of contractive state measurement (CSM) has been discussed by Yuen~\cite{Yuen}, and a concrete mathematical model of the CSM has been proposed by Ozawa~\cite{Ozawa, Ozawa_book}.
Unitary transformation $\hat U$ of CSM proposed by Ozawa is given by Eq.~\eqref{eq:Trans} with $(a,b,c,d)=(0,1,-1,1)$.
We generalize the parameter space of the unitary transformation to $(a,b,c,d)=(0,\lambda^{-1},-\lambda,1)$, where $\lambda$ is a scaling parameter.
The parameters describing the equivalent circuit are then given from Eq.~\eqref{eq:abcd} as $(p,\alpha,\beta,\gamma)=(0,0,-\lambda^{-1},\lambda)$, and corresponding CSM circuit is illustrated in Fig.~\ref{fig:CSM}~\cite{Ozawa2001}.

Suppose that the input state of the CSM circuit is given by $\psi(x)\phi (y)$.
The output state is calculated to be
\begin{equation}
\hat U(0,\lambda^{-1},-\lambda,1) \psi(x) \phi (y)=\psi\left(y/\lambda\right) \phi\left(y-\lambda x\right).
\label{eq:CSM_unitary}
\end{equation}
The probability  of finding an outcome in $a\leq y\leq a+da$ for the probe is given by
\begin{equation}
\begin{split}
&\text{P} \{ a\leq y\leq a+da \} \\
=& \ da \int |\psi (a/\lambda)|^2 |\phi(a- \lambda x)|^2 dx \\
=& \ \frac{1}{\lambda}\left|\psi \left(\frac{a}{\lambda} \right)\right|^2 da.
\label{eq:CSM_prob}
\end{split}
\end{equation}
We note that this probability in Eq.~\eqref{eq:CSM_prob} is independent of probe wave function $\phi$ in contrast to VNM, in which the probability becomes independent of $\phi$ only if $|\phi(y)|^2$ is delta function (see Eq.~\eqref{eq:Neumann_ideal_prob}).
The CSM measurement is therefore a noiseless measurement for an arbitrary state of the probe.

The postmeasurement state of the system, $\psi'_a(x)$, for outcome $y=a$ of the probe is found from Eq.~\eqref{eq:CSM_unitary} to be
\begin{equation}
\psi'_a(x)=\sqrt{\lambda}\phi(a-\lambda x).
\label{eq:CSM_post}
\end{equation}
For the special case of $\lambda=1$, CSM is characterized by measurement operator $\hat M_a=|\phi_a \rangle \langle a|$,
where $|\phi_a \rangle$ is defined by $ \langle x|\phi_{a} \rangle =\phi(a-x)$.
Comparing this with Eq.~\eqref{M_double}, we find close similarity between CSM and the DCNOT measurement.

In respect of VNM, if the position measurement is noiseless, then the postmeasurement state must be a delta function. 
With CSM, in contrast, even if the position measurement is noiseless, the postmeasurement state of the system can be an arbitrary wave function as in Eq.~\eqref{eq:CSM_post}.
However, the postmeasurement state depends on outcome $a$ of the probe.

\subsection{Swapping state measurement}

\begin{figure}[htbp]
\begin{center}
\includegraphics[height=3.3cm,clip]{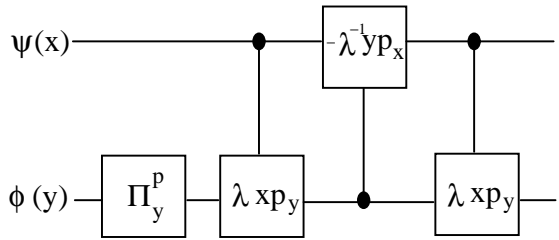}
\caption{Circuit representation of swapping state measurement.}
\label{fig:SSM}
\end{center}
\end{figure}

We consider here a model for quantum measurement which we shall refer to as swapping state measurement (SSM).
Although the notion of SSM has been discussed in Ref.~\cite{Yuen_SSM}, a unitary transformation between the system and the probe that characterizes SSM has not been discussed.
Unitary transformation $\hat U$ of SSM is given by Eq.~\eqref{eq:Trans} with $(a,b,c,d)=(0,\lambda^{-1},(-1)^{p+1} \lambda,0)$, where $\lambda$ is a scaling parameter and $p$ takes on 0 or 1.
It follows from Eq.~\eqref{eq:abcd} that the parameters describing the equivalent circuit are given by $(\alpha,\beta,\gamma)=(\lambda,-\lambda^{-1},\lambda)$, the corresponding SSM circuit being illustrated in Fig.~\ref{fig:SSM}.

Let the input state of SSM be $\psi(x)\phi(y)$.
The output state is then given by
\begin{equation}
\begin{split}
&\hat U(0,\lambda^{-1},(-1)^{p+1} \lambda,0) \psi(x)\phi(y) \\
=&\psi\left( y/\lambda \right)\phi\left( (-1)^{p+1} \lambda x \right).
\end{split}
\label{eq:SSM_unitary}
\end{equation}
We note that the output is not an entangled state but a product state unlike the cases of VNM and CSM.
Comparing the output state with the input one, we find that the SSM circuit exchanges quantum states between the system and the probe, and rescales the coordinate of the system and that of the probe by a factor of $\lambda$ and $\lambda^{-1}$, respectively. 
In addition, the odd (even) parity transformation is applied to the probe for $p=0 \ (p=1)$. 
In the special case of $\lambda=1$ and $p=1$, the output state is given by
\begin{equation}
\hat U(0,1,1,0) \psi(x)\phi(y)=\psi (y )\phi ( x ),
\end{equation}
which is the case in which the initial states of the system and probe are swapped.

The probability of finding an outcome of probe in $a\leq y\leq a+da$ is given from Eq.~\eqref{eq:SSM_unitary} by 
\begin{equation}
\text{P} \{ a\leq y\leq a+da \}=\frac{1}{\lambda}\left|\psi\left(\frac{a}{\lambda}\right)\right|^2 da,
\end{equation}
which is independent of probe wave function $\phi$ as in the case of CSM.
The postmeasurement state of the system is given by
\begin{equation}
\psi'(x)=\sqrt{\lambda}\phi((-1)^{p+1}\lambda x).
\end{equation}
For the special case of $\lambda=1$ and $p=1$, SSM is characterized by measurement operator $\hat M_a=|\phi \rangle \langle a|$.
Comparing this with Eq.~\eqref{M_SWAP}, we find complete parallelism between continuous-variable SSM and its two-level counterpart.
We also note that the postmeasurement system state is independent of the measurement outcome, because the output state is a product state, unlike the cases of VNM and CSM (see Eqs.~\eqref{eq:VNM_post} and \eqref{eq:CSM_post}).
Although we have considered here position measurement, we emphasize that the present noiseless measurement scheme can be applied to an arbitrary observable by using a similar SWAP circuit.

\section{Interaction Hamiltonians and Unitary Gates}
\label{sec:IHUG}

\subsection{Two-Level Systems}

\subsubsection{Composition of Interaction Hamiltonians}

We have shown that the DCNOT and SWAP circuits can be respectively described by using two and three CNOT gates.
In this subsection we construct Hamiltonians describing the individual circuits.

Let us first consider an interaction Hamiltonian of the CNOT circuit.
It is known that the CNOT gate can be described by the following unitary transformation~\cite{Feynman}
\begin{equation}
\hat U^{\text{CNOT}}_{\rm s\text{-} \rm p}=\frac{\hat I+\hat \sigma^{z}_{\rm s}}{2}+\frac{\hat I-\hat \sigma^z_{\rm s}}{2}\hat \sigma^x_{\rm p}.
\end{equation}
Let us consider an operator defined by
\begin{equation}
\hat A\equiv \frac{(\hat I -\hat \sigma^z_{\rm s})(\hat I -\hat \sigma^x_{\rm p})}{4}.
\end{equation}
By noting that operator $\hat A$ is idempotent ($\hat A^2=\hat A$), we can derive the following equation:
\begin{equation}
\exp\left[i \pi \hat A\right]= \hat I-2\hat A=\hat U_{\rm s\text{-} \rm p}^{\text{CNOT}}.
\end{equation}
Therefore, the interaction Hamiltonian of the CNOT circuit is given by
\begin{equation}
\hat H^{\rm CNOT}_{\rm s\text{-}p}=K(\hat I -\hat \sigma^z_{\rm s})(\hat I -\hat \sigma^x_{\rm p}).
\end{equation}

We now consider an interaction Hamiltonian of the DCNOT circuit.
Let us define 
\begin{equation}
\hat B\equiv \frac{\hat \sigma^y_{\rm s}(\hat I-\hat \sigma^x_{\rm p}-\hat \sigma^z_{\rm p})-(\hat I-\hat \sigma^x_{\rm s}-\hat \sigma^z_{\rm s})\hat \sigma^y_{\rm p}}{2\sqrt{3}}.
\end{equation}
Operator $\hat B$ then satisfies $\hat B^3=\hat B$.
As shown in Appendix~\ref{app:DCNOT_Hamiltonian}, we can use this to derive the following relation:
\begin{equation}
\exp\left[ \frac{2i\pi}{3} \hat B \right]= \hat U^{\text{DCNOT}}_{\rm s\text{-} \rm p}.
\label{eq:double_Hamiltonian}
\end{equation}
It follows that an interaction Hamiltonian describing the DCNOT circuit can be expressed by
\begin{equation}
\hat H^{\rm DCNOT}_{\rm s\text{-}p}= K \left[\hat \sigma^y_{\rm s}(\hat I-\hat \sigma^x_{\rm p}-\hat \sigma^z_{\rm p})-(\hat I-\hat \sigma^x_{\rm s}-\hat \sigma^z_{\rm s})\hat \sigma^y_{\rm p} \right].
\label{eq:DCNOT_Hamiltonian}
\end{equation}

The interaction Hamiltonian of the SWAP circuit is known to be an isotropic Heisenberg exchange-interaction Hamiltonian between system spin $\hat{\bm{S}}_{\rm s}$ and probe spin $\hat{\bm{S}}_{\rm p}$~\cite{Loss_DiVincenzo} as
\begin{equation}
\hat H^{\text{SWAP}}\equiv K \left( \hat{\bm{S}}_{\rm s}\cdot \hat{\bm{S}}_{\rm p} \right),
\label{eq:SWAP_Hamiltonian}
\end{equation}
where $\hat{\bm{S}}=\{\hat \sigma^x, \hat \sigma^y, \hat \sigma^z\}$ is a vector of Pauli matrices.
In fact, we can show that the unitary operator defined by
\begin{equation}
\hat U^{\text{SWAP}}=\exp \left[ \frac{i\pi}{4} \hat{\bm{S}}_{\rm s} \cdot \hat{\bm{S}}_{\rm p} \right ],
\end{equation}
has the same effect as that in Eq.~\eqref{USWAP} except for the overall phase factor.

\subsubsection{Representations of unitary circuits in terms of single-qubit rotations and $\left(\hat U^{\text{SWAP}}\right)^{\alpha}$ operators}

It is known that the CNOT circuit can be implemented by using single-qubit rotations and two $\left(\hat U^{\text{SWAP}}\right)^{\alpha}$ operators~\cite{Loss_DiVincenzo,Makhlin,Fan_et_al}, where $\left(\hat U^{\text{SWAP}}\right)^{\alpha}$ is defined by
\begin{equation}
\left(\hat U^{\text{SWAP}}\right)^{\alpha}\equiv\exp\left[\alpha \frac{i\pi}{4} \hat{\bm{S}}_{\rm s} \cdot \hat{\bm{S}}_{\rm p} \right].
\end{equation}
The CNOT circuit can be expressed as
\begin{align}
\hat U^{\text{CNOT}}_{\rm s\text{-} p} 
=& \ \hat U^{\rm H}_{\rm p } \exp \left[\frac{i\pi}{4}\hat \sigma^z_{\rm s}\right] \exp \left[-\frac{i\pi}{4}\hat \sigma^z_{\rm p} \right] \left(\hat U^{\text{SWAP}}\right)^{1/2} \notag \\
&\times  \exp \left[\frac{i\pi}{2}\hat \sigma^z_{\rm s} \right] \left(\hat U^{\text{SWAP}}\right)^{1/2} \hat U^{\rm H}_{\rm p},
\end{align}
where $\hat U^{\rm H}_{\rm p}\equiv(\hat \sigma^x_{\rm p}+\hat \sigma^z_{\rm p})/\sqrt{2}$ is the Hadamard unitary transformation on the probe.

On the other hand, the DCNOT circuit can also be implemented by using single-qubit rotations and two $\left(U^{\text{SWAP}}\right)^{\alpha}$ operators.
The DCNOT circuit can therefore be described by
\begin{align}
\hat U^{\text{DCNOT}}_{\rm s\text{-} \rm p}
=&\ \hat U^{\rm H}_{\rm s}\exp \left[\frac{i\pi}{4}\hat \sigma^z_{\rm s}\right] \exp \left[-\frac{i\pi}{4}\hat \sigma^z_{\rm p} \right] \left(\hat U^{\text{SWAP}}\right)^{1/2}\notag \\
&\times \exp \left[\frac{i\pi}{2}\hat \sigma^z_{\rm s} \right] \left(\hat U^{\text{SWAP}}\right)^{-1/2} \hat U^{\rm H}_{\rm p},
\end{align}
where $\hat U^{\rm H}_{\rm s} \equiv (\hat \sigma^x_{\rm s}+ \hat \sigma^z_{\rm s})/\sqrt{2}$.
The CNOT and DCNOT circuits can therefore be implemented by single-qubit rotations and two $\left(\hat U^{\text{SWAP}}\right)^{\alpha}$ operators.

\subsection{Continuous-Variable Systems}

\subsubsection{Composition of Interaction Hamiltonians}
\label{sec:Composition}

We have discussed that VNM, CSM and SSM can be respectively described by using one, two, and three SUM (or von Neumann) gates.
It is of interest to describe these measurement processes in terms of single interaction Hamiltonians.
The interaction Hamiltonian of VNM is given by Eqs.~\eqref{VN1} and \eqref{VN2} as
\begin{equation}
\hat H^{\rm VNM}= K \hat x \hat p_y.
\label{H_VNM}
\end{equation}

The interaction Hamiltonians of the CSM and the SSM can be obtained by the composition of unitary gates.
It has been shown for real $u,v,w$~\cite{Ozawa1990} that
\begin{equation}
\begin{split}
&\exp\left[\frac{u (\hat x \hat p_x-\hat y\hat p_y) +v \hat y \hat p_x+w \hat x \hat p_y}{i}\right] \psi(x)\phi(y) \\
=&\psi(ax+by)\phi(cx+dy),
\end{split}
\end{equation}
where $D \equiv\sqrt{-(u^2+vw)}$ and 
\begin{equation}
\begin{bmatrix} a & b \\ c & d \end{bmatrix} =\begin{bmatrix} \cos D -u \frac{\sin D}{D} & -v\frac{\sin D}{D} \\
-w \frac{\sin D}{D} & \cos D + u \frac{\sin D}{D} \end{bmatrix}.
\label{eq:Ozawa1990}
\end{equation}

A generalized CSM circuit can be characterized by $(a,b,c,d)=(0,\lambda^{-1},-\lambda,1)$.
From Eq.~\eqref{eq:Ozawa1990}, the parameters of the CSM circuit are given by
\begin{equation}
u=\frac{\pi}{3\sqrt{3}} , \
v=-\frac{2\pi}{3\sqrt{3}} \lambda^{-1}, \
w=\frac{2\pi}{3\sqrt{3}} \lambda.
\end{equation}
The interaction Hamiltonian of CSM is therefore given by 
\begin{equation}
\hat H^{\rm CSM}=K\left[ (\hat x \hat p_x-\hat y \hat p_y)+2\left(\lambda \hat x \hat p_y-\lambda^{-1} \hat y \hat p_x \right) \right].
\label{eq:CSM_Hamiltonian}
\end{equation}
For $\lambda=1$ this Hamiltonian reduces to that obtained by Ozawa~\cite{Ozawa}.

If we ignore parity gate $\hat \Pi_y$, the SSM circuit is characterized by $(a,b,c,d)=(0,\lambda^{-1},-\lambda,0)$.
From Eq.~\eqref{eq:Ozawa1990}, the parameters of the SSM circuit are given by
\begin{equation}
u=0, \
v=-\frac{\pi}{2} \lambda^{-1}, \
w=\frac{\pi}{2} \lambda.
\end{equation}
The interaction Hamiltonian of SSM is therefore given by
\begin{equation}
\hat H^{\rm SSM}_{p=0}=K \left(\lambda \hat x \hat p_y- \lambda^{-1} \hat y \hat p_x \right).
\label{eq:SSM_Hamiltonian}
\end{equation}
For the special case of $\lambda=1$, Eq.~\eqref{eq:SSM_Hamiltonian} reduces to~\cite{D'Ariano}
\begin{equation}
\hat H^{\rm SSM}_{p=0}|_{\lambda=1}=K(\hat x \hat p_y- \hat y \hat p_x).
\label{eq:swap_Hamiltonian}
\end{equation}
This Hamiltonian is proportional to the $z$ component of the angular momentum operator.
Hence, the corresponding unitary evolution amounts to a rotation on the $x-y$ plane, and therefore the time evolution governed by the Hamiltonian in Eq.~\eqref{eq:swap_Hamiltonian} leads to swapping of coordinates $x$ and $y$ after an appropriate interaction time.

On the other hand, if we take parity gate $P_y$ into consideration, the SSM circuit is characterized by $(a,b,c,d)=(0,\lambda^{-1},\lambda,0)$.
As shown in Appendix~\ref{app:SSM}, if we define operator $C$ as
\begin{equation}
\hat C \equiv \frac{\lambda \hat x^2+ \lambda^{-1} \hat p_x^2}{2}+ \frac{\lambda^{-1} \hat y^2+ \lambda \hat p_y^2}{2} -(\hat x\hat y +\hat p_x \hat p_y)-\frac{1}{2},
\end{equation}
we have
\begin{equation}
\exp \left[ \frac{\pi}{2i} \hat C \right] \psi(x) \phi(y) =\psi\left( \lambda^{-1} y \right) \phi \left( \lambda x \right).
\label{eq:SSM_single_Hamiltonian}
\end{equation}
Therefore, the interaction Hamiltonian is given by
\begin{equation}
\hat H^{\rm SSM}_{p=1} = K\hat C.
\label{eq:SSMp1_Hamiltonian}
\end{equation}
For the special case of $\lambda=1$, Eq.~\eqref{eq:SSMp1_Hamiltonian} reduces to
\begin{equation}
\begin{split}
&\hat H^{\rm SSM}_{p=1}|_{\lambda=1} \\
=&K\left[ \frac{ \hat x^2+ \hat p_x^2}{2}+ \frac{ \hat y^2+ \hat p_y^2}{2} -(\hat x\hat y +\hat p_x \hat p_y)-\frac{1}{2} \right].
\end{split}
\end{equation}
We note here that Hamiltonian $\hat H^{\rm SSM}_{p=1}$ with $\lambda=1$ is composed of the two-dimensional harmonic-oscillator Hamiltonians and the interaction Hamiltonian.
Until now, we have assumed that the dynamics between the system and the probe are only governed by the interaction Hamiltonian. 
This assumption implies that the interaction Hamiltonian dominates the free Hamiltonians. 
In the case of Hamiltonian $\hat H^{\rm SSM}_{p=1}$, however, this assumption is no longer valid.

\subsubsection{Representation of unitary circuits in terms of phase-shift, squeezing, and SWAP operators}
\label{sec:Representation}

We have already described continuous-variable measurement processes in terms of von Neumann gates.
However, it appears difficult to experimentally implement the VNM and CSM circuits because the corresponding interaction Hamiltonians~\eqref{H_VNM} and \eqref{eq:CSM_Hamiltonian} are rather artificial.
Here we express the VNM and CSM circuits in terms of the phase-shift operator, two-mode squeezing operator and SWAP operator~\cite{Yurke, Song, Ozawa2003, Braunstein}, all of which are known to be experimentally implemented.

We first show that parity inversion operator $\hat \Pi_x$ can be expressed as
\begin{equation}
\hat \Pi_x \equiv \exp\left[ \frac{\pi}{2i}\left( \hat x^2+ \hat p_x^2-\frac{1}{2}\right) \right].
\label{eq:Parity}
\end{equation}
To see that $\hat \Pi_x$ defined in Eq.~\eqref{eq:Parity} indeed changes the parity of the wave function, i.e.,
\begin{equation}
\hat \Pi_x \psi(x)=\psi(-x),
\end{equation}
we introduce annihilation operator $\hat a\equiv(\hat x+i \hat p_x)/\sqrt{2}$ 
and rewrite Eq.~\eqref{eq:Parity} as
\begin{equation}
\hat \Pi_x=\exp\left[ \frac{\pi}{i} \hat a^{\dagger} \hat a \right].
\label{eq:Pi_x}
\end{equation}
Expanding $\psi(x)$ in terms of the complete set of eigenfunctions of the harmonic oscillator $\varphi_n (x)$ and noting that $\varphi_n (x)$ is an even (odd) function of $x$ for even (odd) $n$, we obtain
\begin{equation}
\psi(x)=\sum_n c_n \varphi_n (x),
\end{equation}
and
\begin{equation}
\begin{split}
\hat \Pi_x \psi(x) &= \sum_n c_n \hat \Pi_x \varphi_n (x) \\
&=\sum_n c_n \exp\left[\frac{\pi}{i}n \right] \varphi_n (x) \\
&=\sum_n c_n (-1)^n \varphi_n (x) \\
&=\sum_n c_n \varphi_n (-x) =\psi(-x).
\end{split}
\end{equation}
Similarly, in terms of $\hat b\equiv (\hat y+ i \hat p_y)/\sqrt{2}$, we obtain 
\begin{equation}
\hat \Pi_y=\exp\left[\frac{\pi}{i} \hat b^{\dagger} \hat b \right].
\label{eq:Pi_y}
\end{equation}
We note that Eqs.~\eqref{eq:Pi_x} and \eqref{eq:Pi_y} describe unitary evolutions implemented with phase shifters.

We next define two-mode squeezing operator $\hat S(r)$ as
\begin{equation}
\hat S(r)\equiv \exp[ir(\hat x \hat p_y+\hat y \hat p_x)],
\end{equation}
which transforms input state $\psi(x)\phi(y)$ into
\begin{equation}
\begin{split}
&\hat S(r)\psi(x)\phi(y) \\
=&\psi(x\cosh r+y\sinh r)\phi(x\sinh r+ y\cosh r).
\end{split}
\label{eq:two_mode_squeeze}
\end{equation}
In terms of operators $\hat{a}$ and $\hat{b}$ already defined, the squeezing operator can be rewritten as
\begin{equation}
\hat S(r)=\exp[r(\hat a \hat b-\hat a^{\dagger} \hat b^{\dagger})],
\end{equation}
which describes the unitary evolution implemented with a non-degenerate parametric amplifier.

Let us also define SWAP operator $\hat T(\theta)$ as
\begin{equation}
\hat T(\theta)\equiv \exp\left[-i\theta(\hat x\hat p_y-\hat y \hat p_x)\right],
\end{equation}
which transforms input state $\psi(x)\phi(y)$ into
\begin{equation}
\begin{split}
&\hat T(\theta)\psi(x) \phi(y) \\
=&\psi(x\cos \theta+y\sin \theta)\phi(-x \sin \theta+y \cos \theta).
\end{split}
\label{eq:rotation}
\end{equation}
In terms of operators $\hat{a}$ and $\hat{b}$, the SWAP operator can be expressed as
\begin{equation}
\hat T(\theta)=\exp[\theta(\hat a \hat b^{\dagger}-\hat b \hat a^{\dagger})],
\end{equation}
which describes the unitary evolution of a beam splitter.

We now consider a unitary circuit as
\begin{equation}
\hat U(r, \theta_1, \theta_2, p)\equiv \hat T(\theta_2 ) \hat S(-r) \hat T(\theta_1) \hat \Pi_y^p,
\end{equation}
where $p$ takes on $0$ or $1$.
It follows from Eqs.~\eqref{eq:two_mode_squeeze} and \eqref{eq:rotation} that $\hat U$ transforms the input state $\psi(x)\phi(y)$ into
\begin{equation}
\hat U(r, \theta_1, \theta_2, p) \psi(x)\phi(y) =\psi(ax+by)\phi(cx+dy),
\label{U2}
\end{equation}
where parameters $a,b,c,d$ are given by
\begin{equation}
\begin{split}
a&=\cosh r\cos(\theta_{1}+\theta_{2})
-\sinh r \sin(\theta_{1}-\theta_{2})
, \\
b&=\cosh r \sin(\theta_{1}+ \theta_{2}) - \sinh r \cos( \theta_{1} - \theta_{2}), \\
c&=(-1)^{p+1} \left[\cosh r \sin (\theta_{1}+\theta_{2})+ \sinh r \cos (\theta_{1}- \theta_{2} \right], \\
d&=(-1)^p \left[\cosh r \cos (\theta_{1}+ \theta_{2})+ \sinh r \sin (\theta_{1}- \theta_{2}) \right].
\end{split}
\label{eq:two_mode_parameter}
\end{equation}
Thus $\hat U(r,\theta_1,\theta_2, p)$ defined by Eq.~\eqref{U2} is equivalent to $\hat U(a,b,c,d)$ defined by Eq.~\eqref{eq:Trans}, provided that Eqs.~\eqref{eq:two_mode_parameter} are met.

We find from Eq.~\eqref{eq:two_mode_parameter} that the VNM circuit with $(a,b,c,d)=(1,0,-\lambda,1)$ can be implemented by using two beam splitters and a nondegenerate parametric amplifier with 
\begin{equation}
r=\ln \left[ \frac{\lambda+\sqrt{\lambda^2+4}}{2}\right], \
\theta_{1}=\theta_{2}=\frac{1}{2}\tan^{-1}\frac{\lambda}{2}, \ p=0.
\label{eq:VNM_twomode_parameter}
\end{equation}
These parameters $r$ and $\theta_1$ satisfy the backaction-evading condition $\sin 2\theta_1=\tanh r$~\cite{Song}.

On the other hand, the CSM circuit with $(a,b,c,d)=(0,\lambda^{-1},-\lambda,1)$ can also be implemented by using two beam splitters and a nondegenerate parametric amplifier with 
\begin{equation}
\begin{split}
r&= \ln \left[ \frac{\sqrt{\Lambda_{+}^{2}+1}+\sqrt{\Lambda_{-}^2+1}}{2} \right],\ p=0, \\
\theta_{1}&= \frac{\tan^{-1} \Lambda_+ -\tan^{-1} \Lambda_-}{2}+ \frac{\pi}{4}, \\
\theta_{2}&= \frac{\tan^{-1} \Lambda_+ +\tan^{-1} \Lambda_-}{2}- \frac{\pi}{4},
\end{split}
\label{eq:CSM_twomode_parameter}
\end{equation}
where $\Lambda_{\pm}\equiv \lambda\pm \lambda^{-1}$.
Ozawa has shown that the CSM circuit for $\lambda=1$ can be constructed by using three $T(\theta)$ and two $S(r)$ operators~\cite{Ozawa2003}. 
In contrast, our method just described can construct the CSM circuit by using two $T(\theta)$ and one $S(r)$ operators. 

SSM circuit $(a,b,c,d)=(0,\lambda^{-1},(-1)^{p+1} \lambda,0)$ can be implemented with
\begin{equation}
r=\ln \lambda,  \ \theta_1= \theta_2 = \frac{\pi}{4}, \ p= 0 \ \text{or} \ 1.
\label{eq:SSM_twomode_parameter}
\end{equation}
For the special case of $\lambda=1$, Eq.~\eqref{eq:VNM_twomode_parameter} becomes
\begin{equation}
r= \ln \varphi, \ \theta_{1}= \theta_{2}= \tan^{-1}\varphi -\frac{\pi}{4}, \ p=0,
\end{equation}
and Eq.~\eqref{eq:CSM_twomode_parameter} becomes
\begin{equation}
r= \ln \varphi, \ \theta_{1}= \tan^{-1}\frac{1}{\varphi} +\frac{\pi}{4}, \ \theta_{2}= \tan^{-1}\frac{1}{\varphi} -\frac{\pi}{4}, \ p=0,
\end{equation}
where $\varphi\equiv(1+\sqrt{5})/2$ is the golden ratio.

The foregoing discussion suggests that, from an experimental point of view, it would be more convenient to use the SWAP operator than the von Neumann operator.
An analogous situation in a spin-1/2 system has been presented in Ref.~\cite{Loss_DiVincenzo}.

As discussed in Appendix~\ref{app:single_mode}, it is also possible to express an arbitrary unitary circuit in terms of the phase shift operator, SWAP operator, and single-mode squeezing operator. This implies that an arbitrary unitary circuit can be implemented using phase shifters, beam splitters and degenerate parametric amplifiers.

\section{Discussions}
\label{sec:Discussions}

\subsection{Three models}
\label{sec:three_models}

We have analyzed noiseless quantum measurements in a two-level system and a continuous-variable system and have shown that there is complete parallelism between the two-level and continuous-variable systems. 
As a consequence, we can categorize those measurements into three classes.

The first class is described by a VNM model.
This model is composed of a CNOT gate or a SUM gate, and the simplest among the three classes. However, it cannot break Heisenberg's noise-disturbance uncertainty relation because the probability 
distribution for the outcome of the measurement depends on the initial state of the probe.

The second class is described by a CSM model.
This model is composed of two CNOT gates or two SUM gates.
This model can break Heisenberg's noise-disturbance uncertainty relation because the probability distribution for the outcome is independent of the initial state of the probe.
The functional form of postmeasurement wave function, $\phi$, is arbitrary; however, outcome $a$ of the measurement does affect it as a shift of the coordinate as shown in Eq.~\eqref{eq:CSM_post}.

The third class is described by an SSM model.
This model is composed of three CNOT gates or three SUM gates.
From the viewpoint of experimental realization, this is the simplest of the three models.
This model also can break Heisenberg's noise-disturbance uncertainty relation.
We can choose the postmeasurement state to be arbitrary and to depend neither on the outcome of the measurement nor on the observable to be measured.
The initialization of the state of the system is therefore straightforward.
Moreover, the SSM circuit does not depend on the observable to be measured, because the SSM circuit exchanges quantum states between the system and the probe.
Therefore, the SSM circuit can be used for noiseless measurement of an arbitrary observable.

\subsection{Scaling parameter}

In the analysis of continuous-variable systems, we have introduced scaling parameter $\lambda$.
The degree of freedom afforded by this parameter can be used to improve the resolution of the probe.

As an example, let us consider the case in which the wave function of the system is sharply localized around two points A and B.
We assume that the probability distribution vanishes at the midpoint between A and B. 
If the resolution of the apparatus is large compared with the distance between A and B,
we cannot distinguish between outcomes A and B from the position measurement.
However, if we introduce scaling parameter $\lambda$ into the measurement,
we can extend the distance between A and B by a factor of $\lambda$.
Therefore, if we choose an appropriate scaling parameter, we can distinguish between outcomes A and B.

\subsection{Gravitational-wave detection}
\label{sec:Gravitation}

The crucial concept in gravitational-wave detection is the standard quantum limit (SQL) for a free-mass position.
If a measurement model does not break the SQL, the model cannot detect gravitational waves.
It is known that CSM for position measurement is an example that breaks the SQL~\cite{Yuen, Ozawa, Ozawa_book}.
We can easily understand, if the wave function of the probe is prepared in a contractive state, SSM also can break the SQL~\cite{Yuen_SSM}.

In respect of CSM, if the measurement outcome of the system is $x=a$, it is necessary to measure in the next measurement the position shift from $x=a$ which is a random variable.
Since the probe for gravitational-wave detection requires a very precise measurement of the position, we should use a probe having fine resolution in a certain range over which the value of the outcome is distributed.
In respect of SSM, however, the postmeasurement system state is independent of the measurement outcome.
Therefore, we only have to use the probe with fine resolution in the vicinity of the origin.

\section{Conclusion}
\label{sec:Conclusion}

In this paper, we have classified noiseless quantum measurements into three classes in both two-level and continuous-variable systems.
We have analyzed the unitary transformation associated with each of these models by using quantum circuits.
We have proposed a model for swapping state measurement (SSM) as a noiseless measurement for an arbitrary observable.
SSM exchanges quantum states between the system and the probe before the projection postulate is applied to the probe.
SSM therefore has the property that the measurement disturbance depends neither on the observable nor on the outcome of the measurement, and consequently the corresponding unitary circuit does not depend on the observable.

\section*{Acknowledgements}
The authors thank K. Murata and S. Nakajima for useful discussions.
This work was supported by a 21st Century COE program at Tokyo Tech, ``Nanometer-Scale Quantum Physics,'' from the Ministry of Education, Culture, Sports, Science and Technology of Japan, and by a CREST program of the JST.

\appendix
\section{Proof of Eq.~(\ref{eq:double_Hamiltonian})}
\label{app:DCNOT_Hamiltonian}
We define an operator $\hat B$ as
\begin{equation}
\hat B\equiv \frac{\hat \sigma^y_{\rm s}(\hat I-\hat \sigma^x_{\rm p}-\hat \sigma^z_{\rm p})-(\hat I- \hat \sigma^x_{\rm s}-\hat \sigma^z_{\rm s})\hat \sigma^y_{\rm p}}{2\sqrt{3}}.
\end{equation}
We use relation $\hat B^3=\hat B$ to show that
\begin{equation}
\begin{split}
\exp [it \hat B]&= \hat I +\hat B \sum_{n=0}^{\infty} \frac{(it)^{2n+1}}{(2n+1)!} +\hat B^2 \sum_{n=1}^{\infty} \frac{(it)^{2n}}{2n!} \\
&=\hat I+(i\sin t) \hat B+ (\cos t -1)\hat B^2.
\end{split}
\end{equation}
By noting that $\hat U^{\rm DCNOT}_{\rm s\text{-}p}$ can be decomposed as $\hat U^{\rm DCNOT}_{\rm s\text{-}p}=\hat U^{\rm CNOT}_{\rm s\text{-}p} \hat U^{\rm CNOT}_{\rm p\text{-}s}$,
we obtain
\begin{equation}
\begin{split}
&\exp\left[\frac{2i\pi}{3} \hat B\right] =I+i\frac{\sqrt{3}}{2}\hat B-\frac{3}{2}\hat B^2 \\
=&\left[\frac{\hat I+\hat \sigma^z_{\rm s}}{2}+\frac{\hat I-\hat \sigma^z_{\rm s}}{2}\hat \sigma^x_{\rm p}\right]\cdot\left[\frac{\hat I+\hat \sigma^z_{\rm p}}{2}+\hat \sigma^x_{\rm s}\frac{\hat I-\hat \sigma^z_{\rm p}}{2}\right] \\
=&\hat U^{\rm CNOT}_{\rm s \text{-}p} \hat U^{\rm CNOT}_{\rm p\text{-}s} \\
=&\hat U^{\rm DCNOT}_{\rm s\text{-}p}.
\end{split}
\end{equation}

\section{Proof of Eq.~(\ref{eq:SSM_single_Hamiltonian}) }
\label{app:SSM}

As shown in Sec.~\ref{sec:IHUG}, parity inversion operator $\hat \Pi_x$ can be expressed as
\begin{equation}
\hat \Pi_x=\exp \left[ \frac{\pi}{2i} \left( \hat x^2+ \hat p_x^2- \frac{1}{2} \right) \right].
\label{eq:parity_inversion}
\end{equation}
We define operator $\hat C$ as 
\begin{equation}
\begin{split}
\hat C=\frac{\lambda \hat x^2+ \lambda^{-1} \hat p_x^2}{2}+ \frac{\lambda^{-1} \hat y^2+ \lambda \hat p_y^2}{2} -(\hat x\hat y +\hat p_x \hat p_y)-\frac{1}{2},
\end{split}
\end{equation}
and consider the following transformation of variables:
\begin{eqnarray}
\hat X \equiv \sqrt{\frac{\lambda}{2}} \hat x- \frac{\hat y}{\sqrt{2\lambda}}, \ \hat P_X \equiv \frac{\hat p_x}{\sqrt{2\lambda}} - \sqrt{\frac{\lambda}{2}} \hat p_y,
\\
\hat Y \equiv \sqrt{\frac{\lambda}{2}} \hat x + \frac{\hat y}{\sqrt{2\lambda}}, \ \hat P_Y \equiv \frac{\hat p_x}{\sqrt{2\lambda}} + \sqrt{\frac{\lambda}{2}} \hat p_y,
\end{eqnarray}
where
\begin{eqnarray}
[\hat X , \hat P_X ] =[\hat Y , \hat P_Y ]=i, \ 
[\hat X , \hat P_Y ] =[\hat Y , \hat P_X ]=0.
\end{eqnarray}
Then $\hat C$ is expressed as 
\begin{equation}
\hat C=\hat X^2 + \hat P_X^2-\frac{1}{2}.
\end{equation}
Hence,
\begin{equation}
\hat \Pi_X=\exp \left[ \frac{\pi}{2i} \hat C \right].
\end{equation}
Therefore,
\begin{equation}
\begin{split}
&\exp \left[ \frac{\pi}{2i} \hat C \right] \psi(x) \phi(y) \\
=&\hat \Pi_X \left[ \psi\left( \frac{Y+X}{\sqrt{2\lambda}} \right) \phi \left(\sqrt{\frac{\lambda}{2}}(Y-X) \right) \right]\\
=&\psi\left( \frac{Y-X}{\sqrt{2\lambda}} \right) \phi \left(\sqrt{\frac{\lambda}{2}}(Y+X) \right) \\
=&\psi\left( \lambda^{-1} y \right) \phi \left( \lambda x \right).
\end{split}
\end{equation}

\section{Representation of a general unitary operator using single-mode squeezing operators}
\label{app:single_mode}

We define single-mode squeezing operators $\hat S_x(r)$ and $\hat S_y(r)$ as
\begin{eqnarray}
& & \hat S_{x}(r)\equiv \exp \left[\frac{ir}{2}(\hat x\hat p_x+\hat p_x \hat x) \right],
\\
& & \hat S_{y}(r)\equiv \exp \left[\frac{ir}{2}(\hat y\hat p_y+\hat p_y \hat y)\right].
\end{eqnarray}
Operator $\hat S_x(r)$ transforms wave function $\psi(x)$ into 
\begin{equation}
\hat S_x(r)\psi(x) =e^{r/2} \psi(e^{r} x).
\label{eq:squeezing}
\end{equation}
In terms of an annihilation operator $\hat a\equiv(\hat x+i \hat p_x)/\sqrt{2}$, $\hat S_x(r)$ can be expressed by $\hat S_x(r)=\exp[\frac{r}{2}( \hat a^2- \hat a^{\dagger 2})]$.
The squeezing operator therefore describes the unitary evolution of a degenerate parametric amplifier.

We now consider a unitary circuit as
\begin{equation}
\hat U(r',\theta_1',\theta_2', p) \equiv \hat T(\theta_2' ) \hat S_y(r')\hat S_x(-r')\hat T(\theta_1') \hat \Pi_y^p.
\end{equation}
It follows from Eqs.~\eqref{eq:rotation} and \eqref{eq:squeezing} 
that $\hat U$ transforms the input state $\psi(x)\phi(y)$ into
\begin{equation}
\hat U(r',\theta_1',\theta_2', p) \psi(x)\phi(y) =\psi(ax+by)\phi(cx+dy),
\end{equation}
where parameters $a,b,c$ and $d$ are given by
\begin{equation}
\begin{split}
a&=\cosh r' \cos(\theta_{1}'+\theta_{2}') - \sinh r'  \cos(\theta_{1}'- \theta_{2}'), \\
b&=\cosh r' \sin(\theta_{1}'+ \theta_{2}') + \sinh r' \sin( \theta_{1}' - \theta_{2}'), \\
c&=(-1)^{p+1}\left[\cosh r' \sin (\theta_{1}'+\theta_{2}')- \sinh r' \sin (\theta_{1}'- \theta_{2}')\right], \\
d&=(-1)^p \left[\cosh r' \cos (\theta_{1}'+ \theta_{2}')+ \sinh r' \cos (\theta_{1}'- \theta_{2}') \right].
\end{split}
\label{eq:single_mode_parameter}
\end{equation}
If we take
\begin{equation}
r' =r, \
\theta_1' =\theta_1-\frac{\pi}{4}, \
\theta_2' =\theta_2+\frac{\pi}{4},
\label{eq:VNM_parameter}
\end{equation}
then Eq.~\eqref{eq:single_mode_parameter} coincides with Eq.~\eqref{eq:two_mode_parameter}.
Therefore, the parameters $r',\theta_{1}',\theta_{2}'$ for the VNM, CSM and SSM circuits can be determined from Eqs.~\eqref{eq:VNM_twomode_parameter}, \eqref{eq:CSM_twomode_parameter} and \eqref{eq:SSM_twomode_parameter}, respectively.

Incidentally, the SSM circuit can also be implemented by using a circuit described by
\begin{equation}
\hat S_y(\ln \lambda^{-1} ) \hat T \left( \frac{\pi}{2} \right) \hat S_y(\ln \lambda ) \hat \Pi_y^p.
\end{equation}
This circuit has the advantage that system's squeezing operator $\hat S_x(r)$ is not necessary.

\end{document}